\documentclass[letter,scriptaddress,twocolumn, prl,showkeys,showpacs,groupedaddress]{revtex4-1}

	\usepackage{amsmath}
	\usepackage{makeidx}
	\usepackage{amsfonts}
	\usepackage{graphicx, xcolor}  
	\usepackage{dcolumn}   
	\usepackage{bm}        
	\usepackage{amssymb}   
	\usepackage[ansinew]{inputenc}
	\usepackage[usenames,dvipsnames]{pstricks}
	\usepackage{subfigure}
	\usepackage{epstopdf}
	\usepackage{pst-grad} 
	\usepackage{pst-plot} 
	\usepackage[colorlinks,hyperindex]{hyperref}
	\hypersetup
	{
		colorlinks,%
		citecolor=black,%
		linkcolor=black,%
		urlcolor=black,%
	}
	\usepackage{auto-pst-pdf}
	
\newcommand{\be}{\begin{equation}}
\newcommand{\ee}{\end{equation}}
\newcommand{\ba}{\begin{array}}
\newcommand{\ea}{\end{array}}

\newcommand{\mg}{\mathcal{G}}

\newcommand{\ms}{\mathcal{S}}

\newcommand{\mi}{\mathcal{I}}

\begin{document}

\title{Evolution of covalent networks under cooling: contrasting the rigidity window and jamming scenarios}
\author{Le Yan, Matthieu Wyart}
\affiliation{Center for Soft Matter Research, Department of Physics, New York University\\ 4 Washington Place, New York, 10003, NY, USA}

\date{\today}

\begin{abstract}
We study the evolution of structural disorder under cooling in supercooled liquids, focusing on  covalent networks. We introduce a model 
for the energy of networks that incorporates weak non-covalent interactions. We show  that at low-temperature, these interactions
considerably affect the network topology near the rigidity transition that occurs as the coordination increases. As a result, this transition  becomes mean-field  and does not present a line of critical points
previously argued for, the  ``rigidity window".   Vibrational modes are then not fractons, but instead are similar to the anomalous modes observed in packings of particles near jamming. 
These results suggest an alternative interpretation for  the intermedia\textcolor{black}{te} phase observed in chalcogenides.

\end{abstract}


\maketitle

The physics of amorphous materials is complicated by the presence of structural disorder, which depends on temperature in supercooled liquids,
and on system preparation in glasses. As a result, various properties of amorphous solids are much less understood than in their crystalline counterparts, such as the non-linear phenomena that control plasticity under stress \cite{Argon79,Falk98} or  the glass transition \cite{Ediger96}, or even linear properties like elasticity.  
Concerning the latter, glasses present a large excess of soft elastic modes,  the so-called boson peak \cite{Phillips81}, and their response to a point perturbation can be heterogeneous on  a scale $l_c$ larger than the particle size \cite{Jaeger96,Leonforte06,Lerner14,Ellenbroek09a}. Recent progress has been made on these questions for short-ranged particles with radial interactions \cite{Liu10}. A central aspect  of these systems is the {\it contact network} made by interacting particles, and its associated average coordination $z$. Scaling behaviors \cite{Liu10} are observed at the unjamming transition where $z\rightarrow z_c$, where $z_c=2d$ is the minimal coordination required for stability \cite{Maxwell64} in spatial dimension $d$. As this bound is approach most of the vibrational spectrum consists of strongly-scattered but extended modes \cite{OHern03a,Silbert05} coined {\it anomalous modes} \cite{Wyart05b}, whose characteristic onset frequency $\omega^*$ vanishes \cite{Silbert05,Wyart05b} and length scale $l_c$ diverges \cite{Silbert05,Lerner14} at threshold. Surprisingly, these critical behaviors can be computed correctly by mean-field approximations, which essentially assume that the spatial fluctuations of coordination are small \cite{Wyart05,Wyart10,DeGiuli14}. Likewise, some detailed aspects of the structure of random close packing are well captured by infinite dimensional calculations  \cite{Charbonneau14,Kurchan13}. However, it is unclear if these results, which assume that structural fluctuations are mild, apply generically to glasses.

\begin{figure}[h!]
\includegraphics[width=0.80\columnwidth]{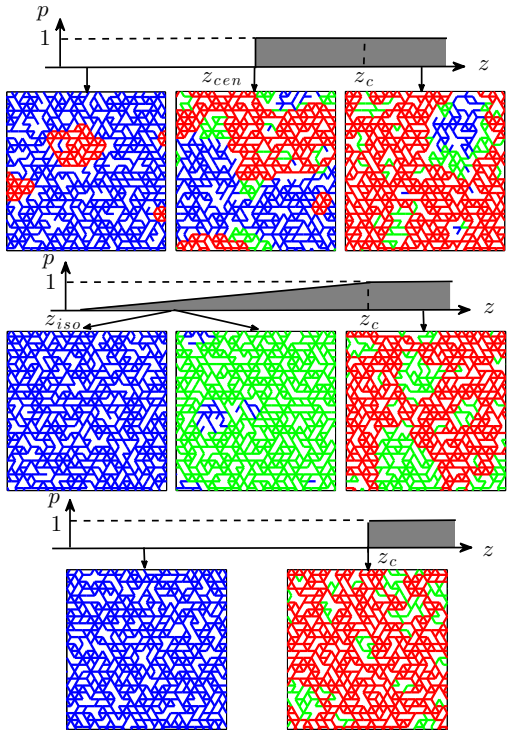}
\caption{\small{Three distinct scenarios for the rigidity transition in chalcogenide glasses. Bonds in blue, green, red corresponds respectively to floppy (under-constrained), isostatic (marginally-constrained)  and over-constrained regions.  $p(z)$ is the probability that a rigid cluster (made of green and red bonds) percolates, as a function of the valence $z$. (a) Rigidity percolation model where bonds are randomly deposited on a lattice. Percolation occurs suddenly and $p(z)$ jumps  from 0 to 1 at $z_{cen}<z_c$. At $z_{cen}$, the rigid network is fractal. (b) The self-organizing network model at  zero temperature. Over-constrained regions are penalized energetically and are absent for $z<z_z$. For $z\in[z_{iso},z_c]$,  $0<p(z)<1$ even in the thermodynamic limit. (c) Mean-field scenario, where $p(z)$ jumps from 0 to 1 at $z_c$, and where the rigid cluster at $z_c$ is not fractal.}}\label{f1}
\end{figure}

In particular, it is generally believed that fluctuations in the structure are fundamental  in covalent glasses such as chalcogenides. 
 In these systems the degree of bonding $z$ plays a role analogous to coordination, and can be changed continuously in compounds such as $Se_xAs_yGe_{1-x-y}$, allowing to go from a polymeric, under-coordinated glass $(x=1,y=0)$ to well-connected structures. Around a mean valence $z_c=2.4$ one expects the network to become rigid \cite{Phillips79,Thorpe85}. Near $z_c$ there is a range of valence, called the intermediate phase \cite{Selvanathan99,Wang00,Georgiev00,Chakravarty05,Wang05}, where the supercooled liquid is strong and the jump of specific heat is small \cite{Tatsumisago90,Bohmer92}, and where the glass almost does not age at all \cite{Selvanathan99,Wang00,Georgiev00,Chakravarty05,Wang05}. Theoretically, at least three distinct scenarios were proposed to describe this rigidity transition, see Fig.~\ref{f1}. Fluctuations are important in the first two. The {\it rigidity percolation} model \cite{Jacobs98,Duxbury99,Feng84,Jacobs95} assumes that bonds are randomly deposited on a lattice. This leads to a second order transition at some $z_{cen}$ where a rigid cluster (a subset of particles with no floppy modes) percolates. Near $z_{cen}$ vibrational modes are fractons \cite{Feng85, Nakayama94}. This model does not take into account that rigid regions cost energy, and thus corresponds to infinite temperature.  To include these effects {\it self-organizing network models} were introduced \cite{Thorpe00,Chubynsky06,Briere07,Micoulaut03}, where rigid regions are penalized. A surprising outcome of these models is the emergence of a rigidity window: a range of valence for which rigidity occurs with a probability $0<p(z)<1$, even in the thermodynamic limit. This rigidity window was proposed to correspond to the intermedia\textcolor{black}{te} phase observed experimentally \cite{Thorpe00}. 
Finally, in the {\it mean-field  or jamming scenario}, fluctuations of coordinations are limited, and $p(z)$ jumps from 0 to 1 at $z_c$. The rigid cluster at $z_c$ is not fractal, and is similar to that of packings of repulsive particles. Specific protocols to generate such networks were used to study elasticity \cite{Wyart08,Ellenbroek09} as well as the thermodynamics and fragility of chalcogenides \cite{Yan13}.

In this Letter we introduce an on-lattice model of networks, and study how  structure and vibrational modes evolve under cooling. Unlike  previous models supporting the existence of a rigidity window \cite{Chubynsky06,Briere07}, our model  includes  weak interactions (such as Van der Waals), always present in addition to covalent bonds. We show numerically and justify theoretically that the rigidity window is not robust: it disappears at low temperature as soon as weak interactions are added. At zero temperature the rigidity transition is then well described by the mean field scenario, and the vibrational modes consist of anomalous modes and not fractons.


\begin{figure}[h!]
\includegraphics[width=1.0\columnwidth]{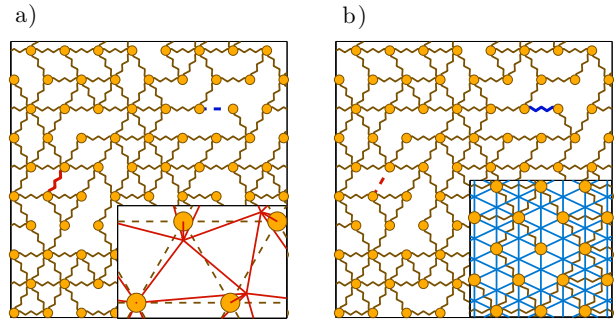}
\caption{\small{Illustration of our model. The triangular lattice is slightly distorted as shown in the inset of (a), and  weak springs connecting all second neighbors are present, as shown in blue in the inset of (b). Our Monte-Carlo considers the motion of strong springs such as that leading from (a) to (b). }}\label{model}
\end{figure}

Our model shares similarity to glasses of polydisperse particles, but it is on-lattice, and  particles are replaced by springs. 
Specifically, in the spirit of \cite{Jacobs95} we consider  a triangular lattice with slight  periodic distortion   to avoid straight lines (non-generic in disordered solids), as shown in Fig.~\ref{model}.
The lattice spacing between neighboring nodes $i$ and $j$ is $r_{\langle i,j\rangle}= 1+\delta_{\langle i,j\rangle}$ where the periodic distortion $\delta_{\langle i,j\rangle}$ is specified in S.I.
Springs of identical stiffness $k$ can jump from an occupied to an unoccupied link, as shown in Fig.~\ref{model}. Their number is controlled by fixing the coordination $z$. The rest length $l_\gamma$ of the spring $\gamma$ positioned on the link $\langle ij\rangle$ is  $l_\gamma=r_{\langle i,j\rangle}+\epsilon_\gamma$, where $\epsilon_\gamma$ is taken from a Gaussian distribution of zero mean and variance $\epsilon^2$ \footnote[1]{The dependence of $l_\gamma$ on link $ij$ is a trick to remove the effect of straight lines on vibrational modes, unphysical for amorphous solids. In an elastic network it could be implemented in two dimensions by forcing the spring to bend in the third dimension, with a position-dependent amount of bending.}.   $k\epsilon^2$ is set to unity as the energy scale. To mimic Van der Waals interactions, we add weak springs of stiffness $k_{\rm w}$ between  second neighbors, so that the coordination of weak springs is $z_{\rm w}=6$. For a given choice of spring location, indicated as  $\Gamma\equiv\{ \gamma\leftrightarrow \langle i,j\rangle\}$, forces are unbalanced if the positions of the nodes are fixed. Instead we allow the  nodes to relax to a minimum of elastic energy $H(\Gamma)$, which depends only on the location of the springs $\Gamma$:
\begin{multline}
\label{e1}
H(\Gamma)=
\min_{\{ {\vec R}_i\}} \sum_\gamma \frac{k}{2} \left[||{\vec R}_i-{\vec R}_j||-l_\gamma\right]^2\\+\sum_{\langle i,j\rangle_2}\frac{k_{\rm w}}{2}\left[||{\vec R}_i-{\vec R}_j||-\sqrt{3}\right]^2
\end{multline}
where ${\vec R}_i$ is the position of node $i$ and ${\langle i,j\rangle_2}$ labels second neighbors. How the minimization of Eq.(\ref{e1}) is performed in practice is described in S.I. 
Having defined an energy functional on all possible network structures $\Gamma$, we perform a Monte Carlo simulation using Glauber dynamics (illustrated in Fig.~\ref{model}) at temperature $T$.


 Our model has two parameters: the temperature $T$ and $\alpha\equiv(z_{\rm w}/d)(k_{\rm w}/k)$  characterizing the relative strength of the weak forces,
estimated from experiments to be of order $\alpha=0.03$ \cite{Yan13}.  We find that we can equilibrate networks in the vicinity of the rigidity transition for $T\geq \alpha$. As we shall see below, for $T\gg1$ we naturally recover rigidity percolation. When $\alpha=0$ and $T\ll1$, a rigidity window appears, as previously reported in \cite{Jacobs95,Thorpe00,Barre05,Chubynsky06}, 
which we exemplify below using $T=3\times10^{-4}$ and $\alpha=0$.  We refer to this condition as {\it strong-force regime}. Finally, our main contention is that for $\alpha>0$ and for $T\leq\alpha$,
the rigidity window disappears, and the rigidity transition is mean-field.  We show that this is already the case  for extremely weak additional interactions $\alpha=T=0.0003$, a condition we refer to as {\it weak-force regime}.

\begin{figure}[h!]
\includegraphics[width=1\columnwidth]{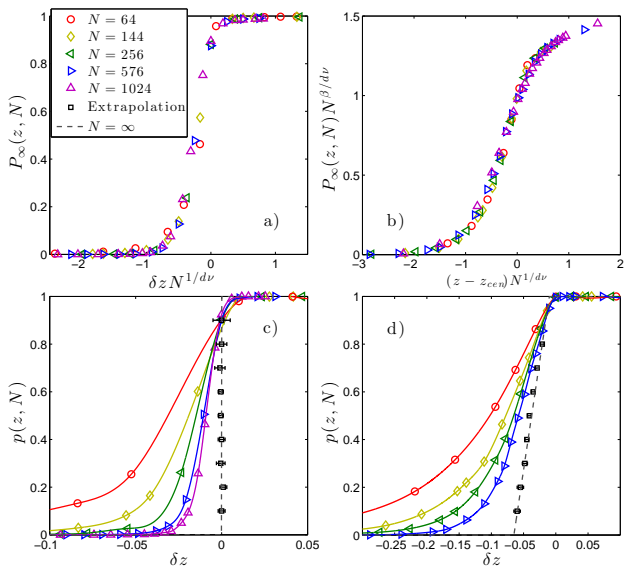}
\caption{\small{$P_{\infty}$ {\it vs} $(z-z_c)N^{1/d\nu}$ in the  {\it weak-force} condition (a) and for $T=\infty$ (b). $p$ {\it vs} $\delta z\equiv z-z_c$ in the {\it weak-force} (c) and {\it strong-force} (d) conditions.  The black squares are extrapolations of the finite $N$ spline curves,  as detailed in the main text.  In (c), the gray line is a step function at $z=z_c$, whereas in (d) it corresponds to the result of~\cite{Chubynsky06}.}}\label{pzsc}
\end{figure}

{\it Percolation probabilities:} the probability $p(z)$ to have a rigid cluster spanning the system, and the  probability $P_{\infty}(z)$ for a bond to belong to this cluster are key quantities to distinguish scenarios. They can be computed for the network of strong springs using the Pebble Game algorithm~\cite{Jacobs97}. For rigidity percolation and infinite system size $N\rightarrow \infty$, $P_{\infty}(z)\sim (z-z_{cen})^{\beta}$. For finite $N$ one then expects \cite{JohnL.Cardy88}  $P_{\infty}(z,N)= N^{-\beta/d\nu}f_{RP}((z-z_{cen})N^{1/d\nu})$, where $f_{RP}$ is a scaling function and $\nu$ the length scale exponent. As shown in Fig.~\ref{pzsc}(b), we recover this result for $T=\infty$, with $z_{cen}=z_c-0.04$, $\beta=0.17$ and $\nu=1.3$, which perfectly matches previous works~\cite{Jacobs96}. Here the Maxwell threshold is set to $z_c=4-6/N$, as expected in two dimensions with periodic boundary conditions. In mean-field, the transition is discontinuous at $z_c$ and one therefore expects $P_{\infty}(z,N)= f_{J}((z-z_c)N^{1/d\nu})$. Our first key evidence that the  {\it weak-force regime} is mean-field is shown in Fig.~\ref{pzsc}(a), where this collapse is satisfied with $\nu=1.0$ - an exponent  consistent with the prediction of ~\cite{Wyart05}.

 Our second key evidence considers $p(z)$, which varies continuously ~\cite{Thorpe00, Chubynsky06,Briere07} in  the rigidity window scenario, but abruptly in mean-field, see Fig.~\ref{f1}. For finite size systems, it turns out to be easier to extract the inverse function $z(p)$, proceeding as follows. 
We first compute $p(z, N)$ for various $z$ and $N$. For each $N$ we use a spline interpolation to obtain continuous curves, as shown in Fig.~\ref{pzsc}(c,d).
We then extract $z(p)$ by fitting the following correction to scaling  $|z(p)-z(p,N)|\sim N^{-1/d\nu}$. Our central result is that for the {\it weak-force regime}, $p(z)$ discontinuously jumps from 0 to 1 at $z_c$ (which simply corresponds to the crossing of the spline lines) as shown in Fig.~\ref{pzsc}(c), again supporting that the mean-field scenario applies. By contrast, in the {\it strong force regime} this procedure predicts a rigidity window   for $z\in [z_c-0.06,z_c]$. This result is essentially identical to previous work using much larger $N$ ~\cite{Chubynsky06} (which is impossible in our model).

\begin{figure}[h!]
\includegraphics[width=0.9\columnwidth]{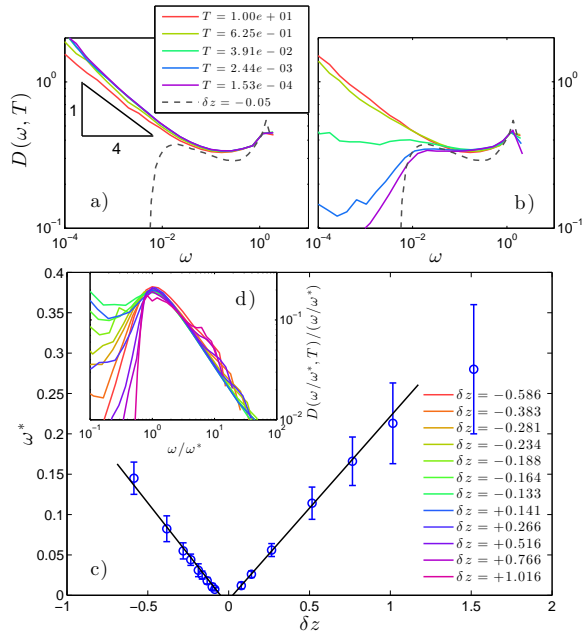}
\caption{\small{ $D(\omega)$ at $z-z_c=-0.05$ and various $T$ indicated in legend for (a) $\alpha=0$ and (b) $\alpha=0.0003$. Gray dashed lines are numerical solution of mean-field networks generated in \cite{Wyart08}. (c) Boson peak frequency $\omega^*$ {\it vs} coordination $z$ for the {\it weak-interaction } regime. $\omega^*$ is defined as the peak frequency of $D(\omega)/\omega^{d-1}$, a quantity shown in (d).}}\label{dos}
\end{figure}

{\it Density of vibrational modes (DOS) $D(\omega,T)$:} 
The DOS is a sensitive observable to characterize  network structure. In the mean-field scenario, anomalous modes appear above a frequency  $\omega^*\sim |z-z_c|$~\cite{Wyart05,During13}, above which the DOS displays a plateau: $D(\omega)\sim \omega^0$, as observed in packings \cite{Silbert05}.
By contrast, at rigidity percolation the rigid cluster is fractal and the spectrum consist of fractons, leading to  $D(\omega)\sim\omega^{\tilde{d}-1}$ ~\cite{Feng85, Nakayama94},  where $\tilde{d}$ is the fracton dimension. Numerically we compute the DOS associated with the  network of strong springs by diagonalization of the stiffness matrix. Within the rigidity window, we find that the DOS is insensitive to temperature for $\alpha=0$ as shown in Fig.~\ref{dos}(a), supporting that normal modes are fractons in the rigidity window, with $\textcolor{black}{\tilde{d}}\approx 0.75$~\cite{Nakayama94}. By contrast, already at small $\alpha=0.0003$, a key observation is that the DOS evolves under cooling  toward the mean-field prediction, as illustrated in Fig.~\ref{dos}(b). At low-temperature, one recovers a frequency scale $\omega^*\sim |z-z_c|$ as shown in Fig.~\ref{dos}(c,d), supporting further that the mean-field scenario applies. Note that there is a very narrow region around $z_c$ where the mean-field prediction does not work well and instead one finds $\omega^*\approx 0$ (see discussion  below).

\begin{figure}[h!]
\includegraphics[width=0.8\columnwidth]{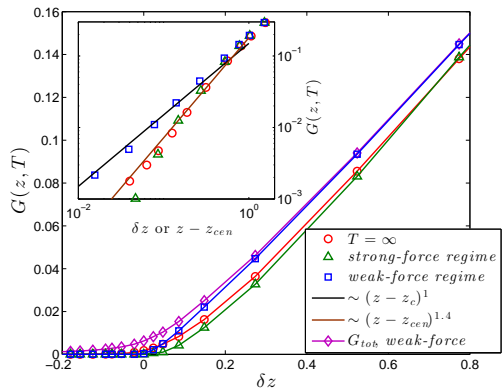}
\caption{\small{Shear modulus of the strong network $G$ {\it vs} $\delta z\equiv z-z_c$ for parameters indicated in legend. The total shear modulus $G_{tot}$ including the effect of weak springs is represented for the {\it weak-interaction regime}.  Inset: same plot in log-log scale, the horizontal axis is $z-z_c$ for low-temperatures conditions (blue and green), and $z-z_{cen}$ at  $T=\infty$ (red).}}\label{Gz}
\end{figure}

{\it Shear modulus $G(z)$:} Lastly, we compute the shear modulus for the strong network numerically, as shown in Fig.~\ref{Gz}. As expected, we find for $T=\infty$ the rigidity percolation result $G(z)\sim(z-z_{cen})^f$,  with $f\approx1.4$~\cite{Broedersz11}. In the {\it weak-interaction regime}, we find that the mean-field result \cite{Thorpe85} $G(z)\sim\delta z$ holds, supporting further our main claim. 
 In the {strong interaction regime}, we find that the shear modulus is zero up to $z_c$. However no power law scaling is found near $z_c$, and $G$ is much lower than in mean-field, in agreement once again with previous models that observed a window~\cite{Chubynsky03}.
%


We have shown numerically that the {\it weak interaction regime} is well-described by the mean-field scenario. To explain this fact,
we argue that this scenario is stable  if $\alpha>0$, but unstable if $\alpha=0$. Consider  the elastic energy per unit volume $E_p$. Qualitatively this quantity is expected to behave as $E_p\sim G_{tot}\epsilon^2$,
where $G_{tot}$ is the total shear modulus  that includes weak interactions, also shown in Fig.~\ref{Gz}. 
The central point is that in mean-field, if $\alpha=0$ then $E_p(z)$ linearly grows for $z>z_c$ and is strictly 0 for $z<z_c$.
It implies that there is no penalty for increasing spatial fluctuations of coordination as long as $z<z_c$ locally. Thus large fluctuations of coordination are expected, 
the mean-field scenario is not stable and one finds a rigidity window instead. 
By contrast, $E_p(z)$ is strictly convex for all $z$ as soon as $\alpha>0$. Then spatial fluctuations of coordination are penalized energetically, and they disappear at low $T$.
In S.I., Fig.~S2, we find numerically that in our model, fluctuations of coordination indeed decay under cooling only if $\alpha>0$.


It is apparent from Fig.~S2 that this process of homogenization is already playing a role at temperatures of order $T\sim 10 \alpha$. 
In practice the glass transition $T_g$ is of order $\alpha$ (the typical covalent bond  energy is between 1 and 10ev, Van der Waals interactions are of between 0.01 and 0.1ev, and the glass transition temperature $T_g$ is of order 100 to 1000K, which is about 0.01 to 0.1ev), supporting that spatial fluctuations of coordination are strongly tamed due to the presence of weak interactions in real covalent glasses.

To conclude, we have argued that weak interactions induce a finite cost to spatial fluctuations of coordination,
which therefore vanish with temperature.   As a consequence, the rigidity transition is mean-field in character,
 if equilibrium can be achieved up to $T=0$. In this light, the mean-field scenario is a convenient starting point to 
 describe these materials. 
 
However, as $T$ increases  fluctuations must be included in the description, as appears in Fig.~\ref{dos}(b).
 Since covalent networks freeze at some  $T_g>0$,  one still expects a finite amount of fluctuations in the glass phase.
Indeed one must cross-over from rigidity percolation at $T=\infty$ to a mean-field scenario at $T=0$. We shall investigate this cross-over in detail elsewhere,
and instead speculate on its nature here. We expect
this cross-over to be continuous, implying  that at any finite temperature,
there is a narrow region around $z_c$ where fluctuations still play a role, and where the transition lies in 
the  rigidity percolation universality class. The size of this region vanishes with vanishing temperature but is finite at $T_g$. Inside this region, one expects  the boson peak to be dominated by fractons, whereas outside 
the mean-field approximation holds and anomalous modes dominate the spectrum. Fig.~\ref{dos}(c) supports this view since already at the very low-temperature considered, there is a narrow region for which $\omega^*\approx 0$, at odds with the mean-field prediction.  As expected, this effect is stronger as $T_g$ increases (as occurs in our model when $\alpha$ increases), as illustrated in Fig.~S3 of S.I.  This qualitative difference in elasticity 
\textcolor{black}{is} likely to affect thermodynamic and aging properties near the glass transition, since these properties are known to be strongly coupled \cite{Novikov05,Yan13}.
The region surrounding the rigidity transition where fluctuations are important is thus a natural candidate for the intermedia\textcolor{black}{te} phase observed in chalcogenides, which would then  result from a dynamical effect, namely the freezing of fluctuations at the glass transition.

\begin{acknowledgements}
We thank E.~Lerner, E.~DeGiuli, G.~D\"uring for discussions, J.~Lin for  discussions leading to our finite size extrapolation method, and D.~Jacobs for sharing the pebble game code. This work has been supported primar-
ily by the National Science Foundation CBET-1236378, and partially by the Sloan
Fellowship, the NSF DMR-1105387, and the Petroleum Research Fund 52031-DNI9.
\end{acknowledgements}

\bibliography{Wyartbibnew}
\begin{appendix}
\renewcommand{\theequation}{S\arabic{equation}}
\setcounter{equation}{0}
\setcounter{figure}{0}
\renewcommand{\thefigure}{S\arabic{figure}}
\section{Supplementary Information}
\subsection{A. Periodic distortion of triangular lattice.}
\begin{figure}[h!]
   {\def\svgwidth{0.48\columnwidth}
   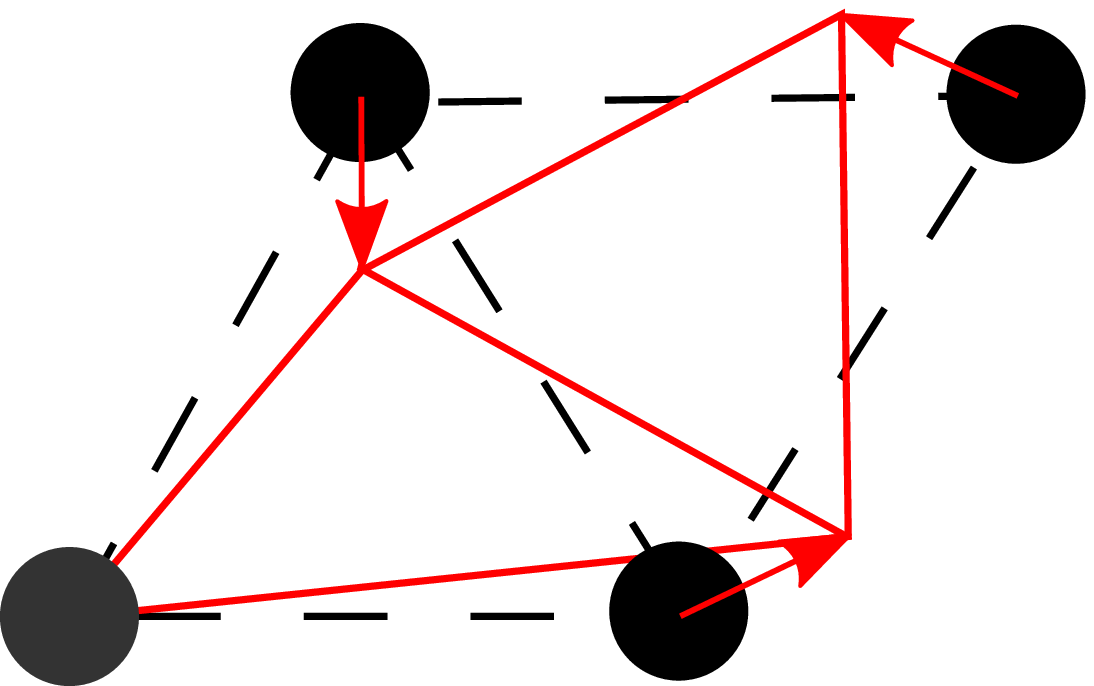}\hfill
   {\includegraphics[width=0.48\columnwidth]{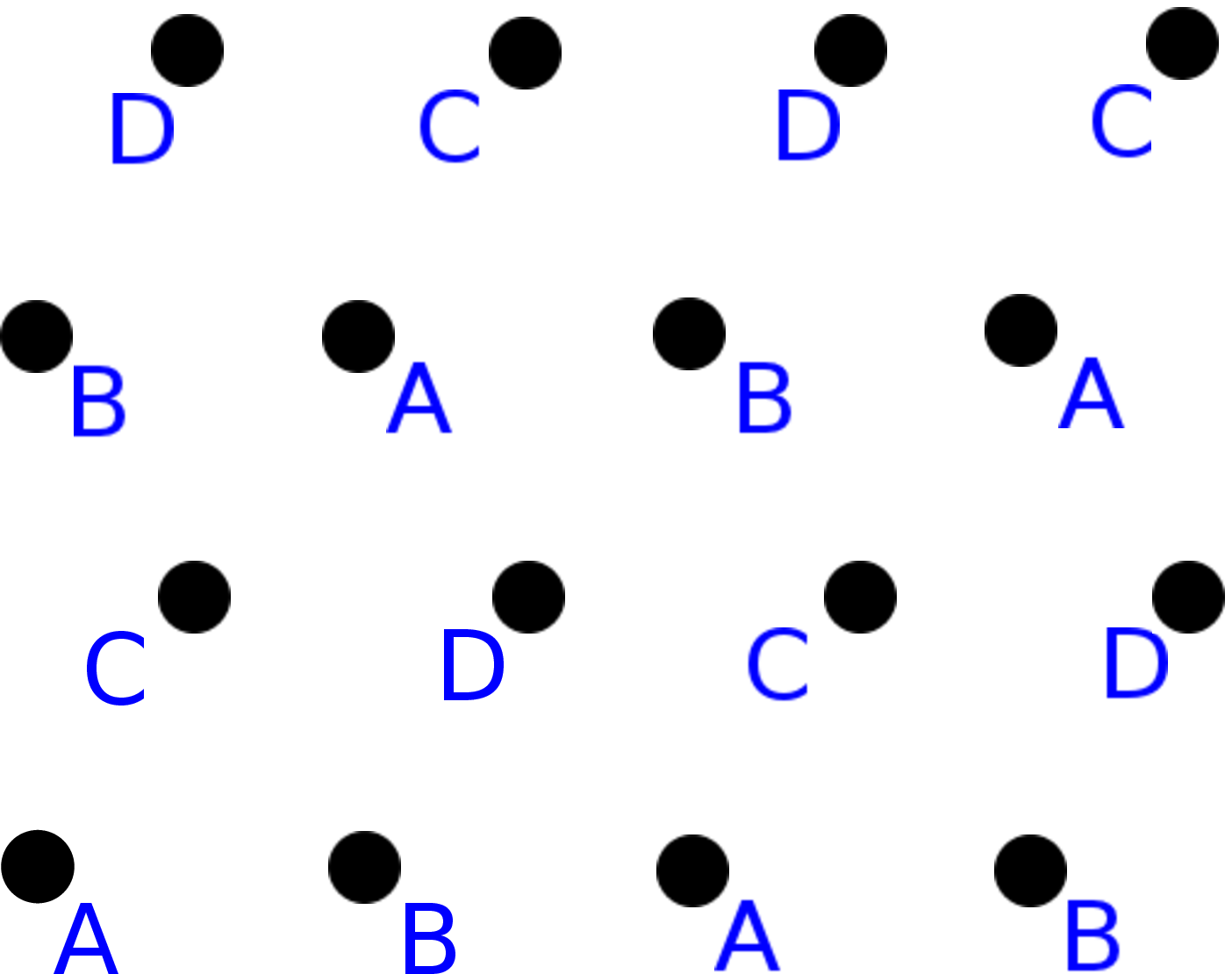}}\\
 \caption{Illustration of distortion of the triangular lattice, performed to remove straight lines.}
 \label{crystalline}
\end{figure}

In our model, we introduce a slight distortion of the lattice to remove the straight lines that occur in a triangular lattice, in the  spirit of \cite{Jacobs95}. Such straight lines would lead to unphysical localized floppy modes orthogonal to the lines.  In ~\cite{Jacobs95} random disorder is introduced to achieve this goal. 
Instead, we seek to distort the lines while avoiding frozen disorder (the only disorder we use corresponds to the polydispersity of the spring rest length, but it does not break translational symmetry  because springs can move). We group nodes by four, labeled as A B C D in Fig.~\ref{crystalline}. \textcolor{black}{One group forms a cell }
of our crystalline lattice. Each cell is distorted identically as follows: node A stays in place,  while nodes B, C, and D move by some distance $\delta$: B along the direction perpendicular to BC, C along the direction perpendicular to CD, and D along the direction perpendicular to DB, as illustrated in the figure.  $\delta$ is set to  $0.2$. 

\subsection{B. Numerical computation of the elastic energy.}
The energy $H(\Gamma)$ of a given spring configuration $\Gamma\equiv\{\gamma\leftrightarrow\langle i,j\rangle\}$ is defined in Eq.(1) of the main text as a minimization on the position\textcolor{black}{s} of the nodes. This minimum can be calculated using conjugate gradient methods. However for small mismatches $\epsilon$, it is more efficient to use  linear algebra  ~\cite{Yan13}, as we now recall.
Consider a displacement field $\delta\vec{R}_i\equiv\vec{R}_i-\vec{R}_{i0}$, where $\vec{R}_{i0}$ is the position of the node $i$ in the crystal described in the previous section. We define the distance $||\vec{R}_{i0}-\vec{R}_{j0}||\equiv r_{\langle i,j\rangle}$. At first order in $\delta\vec{R}_i$, the distance among  neighboring nodes can be written as:
\be
||\vec{R}_i-\vec{R}_j||=r_{\langle i,j\rangle}+\sum_{k}\ms_{\langle i,j\rangle,k}\delta\vec{R}_k+o(\delta\vec{R}^2)
\label{dist}
\ee
Where $\ms$ is the structure matrix, which gives the linear relation between displacements and changes of  distances, as indicated in Eq.(\ref{dist}). Minimizing Eq.(1) in the main text leads to \cite{Yan13}:
\be
H(\Gamma)=\frac{k}{2}\sum_{\gamma,\rho}\epsilon_{\gamma}\mg_{\gamma,\rho}\epsilon_{\rho}+o(\epsilon^3)
\label{Hamiltonian}
\ee
where $\mg=\ms(\ms^t\ms+\alpha\mi)^{-1}\ms^t$, and $\bullet^t$ is our notation for the transpose of a matrix.
In practice, we solve Eq.(\ref{Hamiltonian}) for every configuration $\Gamma$ our Monte Carlo considers. One issue with Eq.(\ref{Hamiltonian}) is that the inverse in the expression for $\mg$ is ill-defined when $\alpha=0$  if floppy modes are present in the network. To study the  case $\alpha=0$, we implement the Pebble Game algorithm~\cite{Jacobs95,Jacobs97} to distinguish stressed, hyperstatic clusters from floppy or isostatic regions.  Since only the stressed regions can contribute to the energy, we  reduce the matrix $\ms$ to this associated subspace, and solve Eq.(\ref{Hamiltonian}) in this subspace.  We have compared this method and a direct minimization via conjugate gradients; the two results coincide within $1\%$ as long as $\epsilon\lesssim0.01$. In the main text our results are based on Eq.(\ref{Hamiltonian}), and thus hold as long as $\epsilon$ is small enough. In this case the choice of $\epsilon$ only affects the energy scale.

\begin{figure*}[T]
\includegraphics[width=1.0\textwidth]{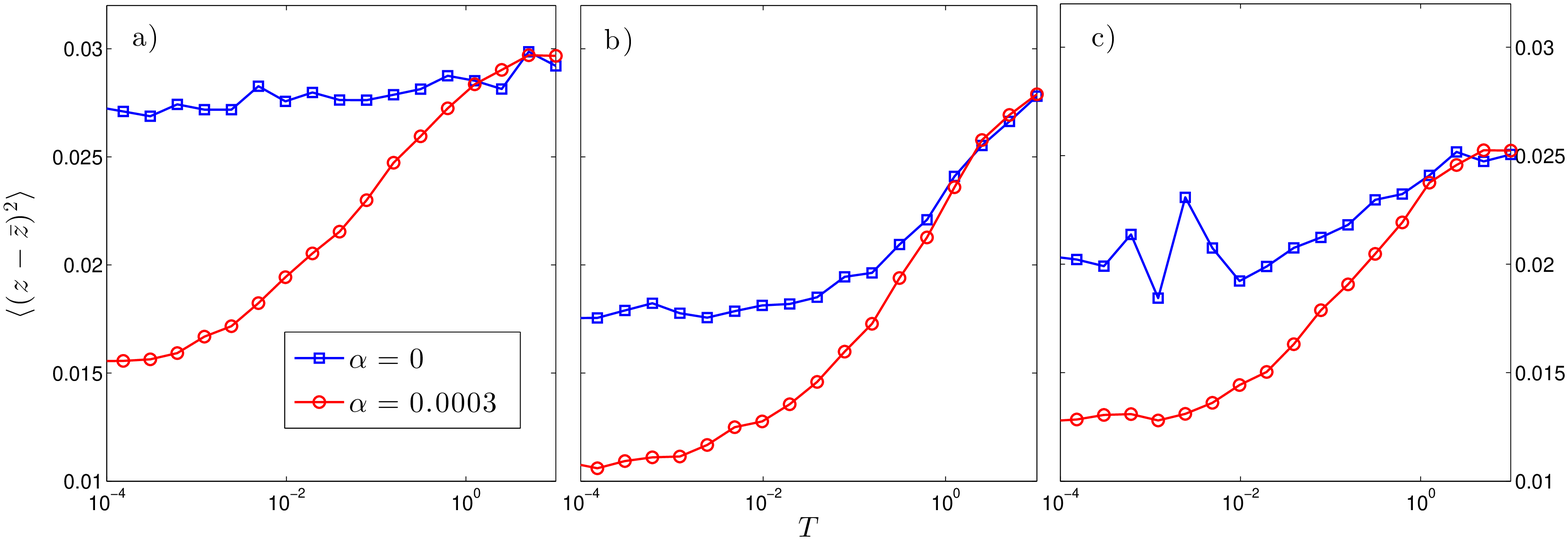}
\caption{\small{Fluctuations of coordination $\langle(z-\bar{z})^2\rangle$ {\it vs} temperature $T$ for different  $\alpha$ as indicated in legend. The network size is $N=256$ and the block size is $N^*=64$. Mean coordination number corresponds to a) $\bar{z}-z_c=-0.383$, b) $\bar{z}-z_c=-0.055$, c) $\bar{z}-z_c=0.523$.}}\label{zfT}
\end{figure*}
\subsection{C. Removal of fluctuations under cooling: numerical evidence}
The mean coordination number of the whole network is fixed in our model; in this section we denote it as $\bar{z}$. To characterize spatial fluctuations of coordination, we divide the network into four identical blocks of size $N^*=N/4$ sites. We then measure the coordination number $z$ in each  bloc\textcolor{black}{k}, and in many configurations equilibrated at some temperature $T$. We then compute the variance $\langle(z-\bar{z})^2\rangle$, where the average is over all blocks and configurations. Fig.~\ref{zfT} shows this quantity versus temperature for three choices of excess coordination $\delta z=\bar{z}-z_c$, corresponding to  a) below c) above  and b) near the rigidity transition. For all these choices we find that the amplitude of fluctuations does not vary for low temperatures when  $\alpha=0$. By contrast when $\alpha>0$, fluctuations of coordination are smaller  at low temperature, where they continue to decay under cooling.

\subsection{D. Effect  of frozen-in fluctuations of coordination at $T_g$}

\begin{figure}[H]
\includegraphics[width=1.\columnwidth]{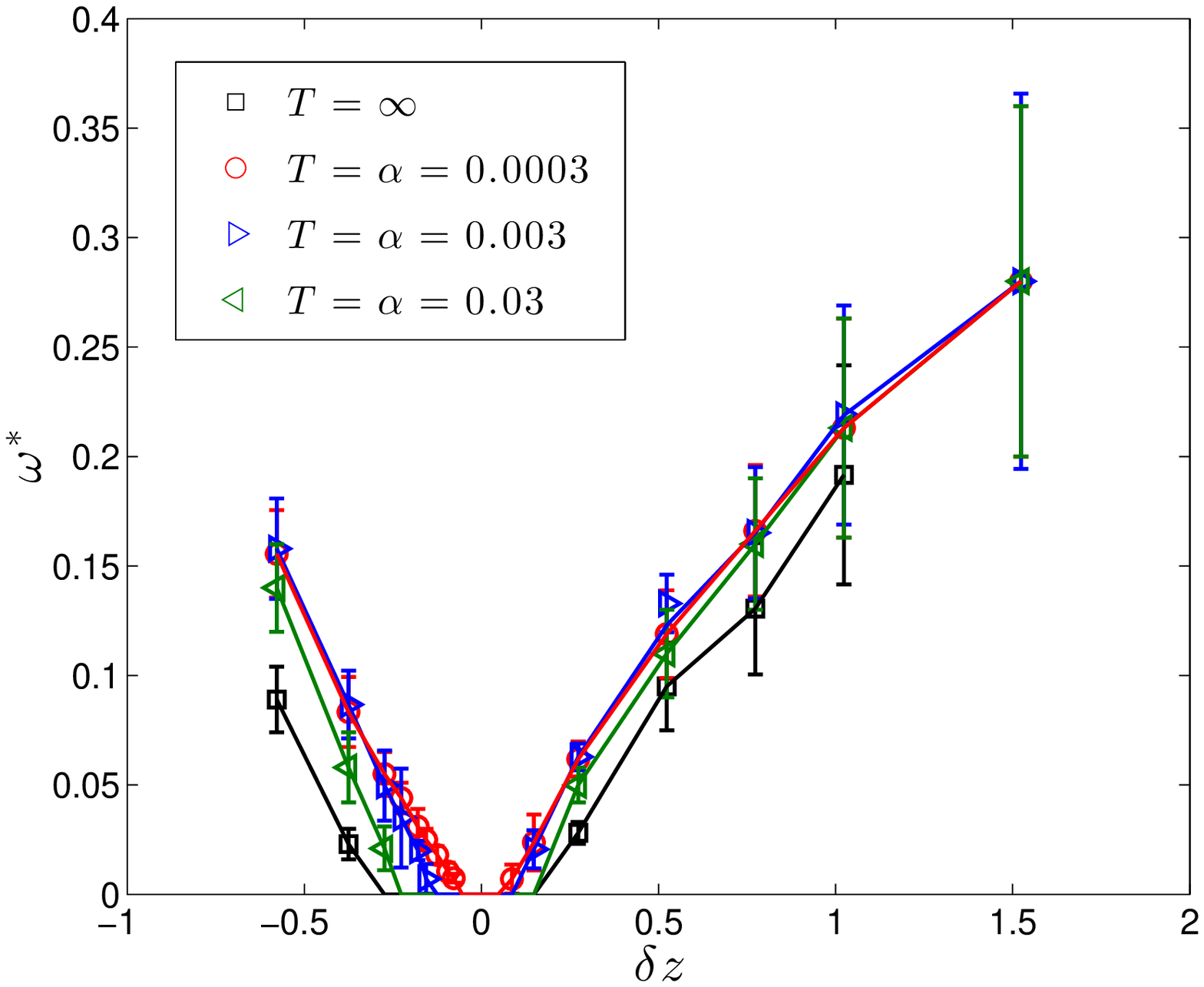}
\caption{\small{Boson peak frequency $\omega^*$ as  a function of excess coordination $\delta z=z-z_c$ for different $\alpha$  as indicated in legend, at temperatures $T=\alpha$. $\omega^*=0$ indicates that no maximum was observed in $D(\omega)/\omega^{d-1}$, consistent with the presence of fractons at very low frequency.}}\label{flu}
\end{figure}

To study the role of frozen-in spatial fluctuations of coordination, we increase $T_g$ in our model, which can be achieved  by increasing the strength of weak interactions $\alpha$. 
We can equilibrate our system up to temperatures of order $T= \alpha$, and in what follows we fix these two parameters to be equal. We then study the vibrational properties of the network of strong springs by computing the boson peak frequency $\omega^*$, defined as in the main text as the maximum of $D(\omega)/\omega^{d-1}$. If we observe no maximum in this quantity we posit that $\omega^*=0$. Results are shown in Fig.~\ref{flu}. The key point is that as $T_g$ increases, a broader region appears in the vicinity of $z_c$ where mean-field predictions do not apply. Instead one finds that for a range of coordination, $\omega^*\approx 0$, consistent with the presence of fractons.  

\end{appendix}

\end{document}